\documentclass{emulateapj}
\slugcomment{Accepted for publication in the Astrophysical Journal}

\shorttitle{Strong Lensing by Subhalos: Detection Probabilities}
\shortauthors{Riehm et al.}
 
\begin{document}

\title{Strong Lensing by Subhalos in the Dwarf-Galaxy Mass Range II: Detection Probabilities}

\author{T. Riehm\altaffilmark{1}, E. Zackrisson\altaffilmark{2,1,3}, E. M\"ortsell\altaffilmark{4} \& K. Wiik\altaffilmark{2}}
\email{teresa@astro.su.se}

\altaffiltext{1}{Department of Astronomy, Stockholm University, AlbaNova University Center, SE-106~91 Stockholm, Sweden}
\altaffiltext{2}{Tuorla Observatory, Department of Physics and Astronomy, University of Turku, V\"ais\"al\"antie 20, FI-215~00 Piikki\"o, Finland}
\altaffiltext{3}{Department of Astronomy and Space Physics, Box 515, SE-751 20 Uppsala, Sweden}
\altaffiltext{4}{Department of Physics, Stockholm University, AlbaNova University Center, SE-106 91 Stockholm, Sweden}

\begin{abstract}
The dark halo substructures predicted by current cold dark matter simulations may in principle be detectable through strong-lensing image splitting of quasars on small angular scales (0.01 arcseconds or below). Here, we estimate the overall probabilities for lensing by substructures in a host halo closely aligned to the line of sight to a background quasar. Under the assumption that the quasar can be approximated as a point source, the optical depth for strong gravitational lensing by subhalos typically turns out to be very small ($\tau <$ 0.01), contrary to previous claims. We therefore conclude that it is currently not feasible to use this strategy to put the simulation predictions for the dark matter subhalo population to the test. However, if one assumes the source to be spatially extended, as is the case for a quasar observed at radio wavelengths, there is a reasonable probability for witnessing substructure lensing effects even at rather large projected distances from the host galaxy, provided that the angular resolution is sufficient. While multiply-imaged, radio-loud quasars would be the best targets for unambiguously detecting dark matter subhalos, even singly-imaged radio quasars might be useful for setting upper limits on the abundance and central surface mass density of subhalos.
\end{abstract}


\section{Introduction}
\noindent
In the standard cold dark matter (CDM) scenario, as well as in several slight modifications thereof, dark matter halos are assembled hierarchically from smaller subunits. At the time of merging, some of these subunits are disrupted and dispersed into the smooth dark matter component of the halo, whereas others temporarily survive in the form of subhalos. A long-standing problem with this picture is that the number of subhalos predicted by CDM simulations is orders of magnitudes higher than the known number of satellite galaxies in the vicinity of the Milky Way \citep[e.g.][]{klypin99,moore99,diemand07b}. There are several possible ways out of this dilemma: either the CDM scenario is incorrect, the simulation predictions are wrong, or the majority of these subhalos must somehow have evaded detection. The latter alternative is quite viable, provided that the baryonic content of these subhalos have been either lost or prevented to form stars \citep{bullock00,benson02,somerville02,kravtsov04,moore06,strigari07}. Recent studies have shown that when including these physical processes together with observational biases most of the predicted Milky Way satellites would lie outside the reach of current surveys \citep[e.g][]{tollerud08,koposov09,maccio09}. If such ''dark galaxies'' do indeed exist, gravitational lensing may offer one of the most promising ways to detect them. 

One tell-tale signature of dark matter subhalos in the 10$^6$--10$^{10}$ M$_{\odot}$ range would be gravitational millilensing, i.e. image splitting at a characteristic separation of milliarcseconds \citep[e.g.][]{wambsganss92,baryshev97,metcalf01,yonehara03}. Based on a null detection of millilensing in a sample of 300 quasars observed with the VLBI, \citet{wilkinson01} demonstrated that the vast majority of quasars do not show any signs of millilensing, and were able to impose an upper limit of $\Omega$ $<$ 0.01 on the cosmological density of point-mass dark matter objects in this mass range. Unfortunately, this limit is insufficient to set any useful constraints on subhalos predicted by CDM, since their lensing properties are very different from those of point-masses \citep[][hereafter paper I]{ez08}. Adopting more realistic subhalo density profiles would significantly raise the above limit. To put the CDM subhalo predictions to the test, it has instead been suggested that one should target quasars which are already known to be gravitationally lensed on arcsecond scales, as one can then be sure that there is a massive halo well-aligned with the line of sight, which substantially increases the probability for subhalo millilensing \citep{yonehara03}. Indeed, the magnification associated with millilensing has long been suspected to be the cause of the flux ratio anomalies seen in such systems \citep[e.g.][]{mao98,kochanek04}. However, current CDM simulations appear to predict too few subhalos to explain the flux ratios of many of these systems \citep[e.g.][]{mao04,metcalf05a,maccio06,xu09}. Thus, other mechanisms like lensing by low-mass field halos \citep{metcalf05b,miranda07}, stellar microlensing in the lens galaxy \citep{schechter02} and absorption or scintillation in the interstellar medium \citep{mittal07} may also be at work. Subhalo millilensing has also been advocated as an explanation for strange bending angles of radio jets \citep{metcalf02} and image positions which smooth halo models seem unable to account for \citep[e.g.][]{koopmans02,biggs04,more08}.

In paper I, we estimated the image separations for the subhalo density profiles favoured by recent N-body simulations, and compared these to the angular resolution of both existing and upcoming observational facilities.
In this second paper in the series, we assess the probability for subhalo millilensing of a quasar as a function of projected radius from the center of a foreground lens galaxy. The paper is structured as follows. In $\S$\ref{section:host}, the properties of host halos are discussed. The subhalo population is described in $\S$\ref{section:subhalo}. In $\S$\ref{section:point}, we compute the optical depth for subhalo lensing of a point-source as a function of projected radius from the host halo. The expected number of intervening subhalos located in a galaxy-sized host halo along the line of sight to an extended source is estimated in $\S$\ref{section:ext}. Finally, in $\S$\ref{section:disc}, we discuss several effects that could affect our predictions and present our conclusions. 

Throughout the paper, we assume a $\Lambda$CDM cosmology with $\Omega_{\Lambda}$ = 0.762, $\Omega_M$ = 0.238 and $h$ = 0.73 ($H_0$ = 100 $h$ km s$^{-1}$ Mpc$^{-1}$) in concordance with the WMAP 3-year data release \citep{spergel07}.

\section{Host halo properties}
\label{section:host}
\noindent
In this study, we focus on host halo lenses in the typical galaxy mass range which spans $10^{10}$ M$_{\odot} \lesssim M \lesssim 2 \times 10^{13}$ M$_{\odot}$ \citep[e.g.][]{li03}. The upper limit is set by the fact that more massive dark matter halos correspond to groups or clusters of galaxies. Although it might be interesting to investigate the dark substructure population in these systems, they are presumably fundamentally different from galaxy-sized halos. This is partly due to a difference in the distribution of the luminous component. Furthermore, it is only for a critical mass $\lesssim 10^{13}$ M$_{\odot}$ that cooling of the baryonic halo component can lead to a concentration of the baryons to the inner parts of the mass profile, resulting in a steep inner density profile \citep{blumenthal86}. Therefore, halos at higher masses are expected to exhibit shallow inner density profiles similar to the profile proposed by \citet[][hereafter NFW]{nfw97}, which has been confirmed by several lensing studies \citep[e.g][]{li02,mandelbaum06}. Most halos with masses below $10^{10}$ M$_{\odot}$ fall in the dwarf galaxy mass range, which corresponds to the subhalo masses we are interested in. A host halo in this mass range would comprise subhalos with even smaller masses (typically lighter than 1\% of the host mass, see below) and therefore not cover the entire relevant subhalo mass range. 
 
Unless stated otherwise, we assume the dark subhalos to reside within a Milky-Way like host halo of mass $M = 1.8 \times 10^{12}$ M$_{\odot}$ at a redshift $z_l$ = 0.5, while the sources are placed at redshift $z_s$ = 2. 

The singular isothermal sphere (SIS), with density profile $\rho \propto r^{-2}$ has proved to be a successful model for luminous galaxy-mass lenses \citep[e.g.][]{rusin03,treu04}. This is believed to be due to the substantial contribution from baryons to the inner regions of these objects, as N-body simulations based on the CDM scenario predict the dark matter halo to be substantially less centrally concentrated (NFW). The mass of an SIS inside a three-dimensional radius $R$ is 
\begin{equation}
M(R) = \frac{2 \sigma_v^2 R}{G}
\end{equation}
with the one-dimensional velocity dispersion $\sigma_v$. 
The projected surface mass density of an SIS is then given by 
\begin{equation}
\Sigma(\xi) = \frac{\sigma_v^2}{2G\xi},
\end{equation}
where $\xi$ is the projected distance from the center of the lens.

The convergence $\kappa$ and the shear $\gamma$ for an SIS are identical and simply proportional to its surface density \citep[e.g.][]{schneider92}:
\begin{equation}
\kappa = \gamma = \frac{\Sigma}{\Sigma_{crit}}
\end{equation}
with the critical surface density defined as
\begin{equation}
\Sigma_{crit} = \frac{c^2}{4\pi G}\frac{D_s}{D_l D_{ls}},
\end{equation}
where $D_s$, $D_l$ and $D_{ls}$ represent the angular diameter distance from the observer to the source, from the observer to the lens, and from the lens to the source, respectively.

The lensing magnification from the host halo $\mu$ is then given by
\begin{equation}
\mu = [(1-\kappa)^2-\gamma^2]^{-1} = (1 - 2 \kappa)^{-1}.
\end{equation}

The host halo is truncated at the virial radius $R_{200}$ defined as the radius at which the mean enclosed density equals
200 times the mean mass density of the Universe at redshift $z$,
\begin{equation}
R_{200} = \left(\frac{3 M}{800 \pi \bar{\rho}}\right)^{1/3}.
\end{equation}

\section{Subhalo populations}
\label{section:subhalo}
\noindent
We are interested in dark subhalos with masses $m$ within the range $10^6 \lesssim m/$M$_\odot \lesssim 10^{10}$. These limits are partly due to the current resolution of the numerical simulations used to predict the existence of substructure. Furthermore, as has been shown in paper I, the lower mass limit roughly corresponds to a minimum halo mass for which the expected image separation could be resolved with either current or upcoming observational facilities. The upper mass limit roughly corresponds to the mass below which the discrepancy between the number densities of luminous galaxies and dark matter halos starts to become severe \citep[e.g.][]{verde02,bosch03}.

We assume a subhalo population following the model proposed by \citet{gao04}. The subhalo abundance per unit halo mass (ignoring the high mass cut-off, $m > 0.01 M$) can be approximated by
\begin{equation}
\frac{dn}{dm} = 10^{-3.2} \left(\frac{m}{h^{-1} M_{\odot}}\right)^{-1.9} h M_{\odot}^{-1} .
\end{equation}

\citet{diemand07a} confirmed this finding and extended it to lower subhalo masses in a recent high resolution simulation of CDM substructure in a Milky Way-sized halo. A fraction $f_{sub}(M,m_{min},m_{max})$ of the mass $M$ of a host halo is in the form of subhalos with minimum and maximum masses $m_{min}$ and $m_{max}$, respectively. Here we assume subhalo masses $m$ in the range $4 \times 10^{6} \leq m/$M$_\odot \leq 10^{10}$, corresponding to the interval probed by \citet{diemand07a}. The total mass fraction of a host halo with $M$ = $1.8 \times 10^{12}$ M$_{\odot}$ in subhalos is then about 5$\%$ for this subhalo mass range.  

Within this simulation, it has also been shown that the subhalo number density profile can be fitted by the following form \citep{madau08}:
\begin{equation}
\label{eq:n}
\frac{n(<x)}{N} = \frac{12x^3}{1+11x^2},
\end{equation}
where $x$ is the distance to the host center in units of $R_{200}$, $n(<x)$ is the number of subhalos within $x$ and $N$ is the total number of subhalos inside $R_{200}$.

\section{Point-sources}
\label{section:point}
\noindent
In the optical wavelength region, a quasar can for the purpose of this paper be approximated by a point source. Under the assumption that the lenses do not overlap along the line of sight, the optical depth $\tau$ represents the fraction of a given patch of the sky that is covered by regions in which a point source will be lensed. 

\begin{equation}
\tau(\xi,\kappa, \gamma) = \frac{1}{S}\int_{m_{min}} ^{m_{max}} \sigma_{lens}(m,\kappa,\gamma) \frac{dn(\xi)}{dm} dm
\end{equation}
with the area of a patch on the sky $S$ and $n(\xi)$ the number of subhalos projected on $S$ at the projected distance $\xi$ from the center of the host halo which has been derived numerically using equation (\ref{eq:n}). Here, $\kappa$ and $\gamma$ are the external convergence and shear induced by the host halo, respectively, and $\sigma_{lens}$ denotes the cross section for a single subhalo lens  within the potential of the host halo. 
In the limit of small $\tau$, the optical depth can therefore directly be used as an estimate of the lensing probability. 

\citet{keeton03} derived an analytic expression for the lensing cross section of an SIS subhalo within a host halo potential, 
\begin{equation}
\sigma_{lens}(m,\delta,\kappa,\gamma) = \mu A(m,\delta,\kappa,\gamma),
\end{equation} 
where $A$ corresponds to the area in the source plane where the total magnification perturbation due to the substructure is stronger than $\delta$. 
In the limit $\left|\delta\right| \rightarrow \infty$, this analysis gives the area inside the caustic where we expect to obtain multiple images from substructure millilensing.
A discussion on the validity of the SIS assumption for substructures can be found in $\S$\ref{section:disc}. The area $A$ can be expressed through the Einstein radius of the subhalo, $b$, which is defined as
\begin{equation}
b = 4 \pi \left(\frac{\sigma_v}{c}\right)^2\frac{D_{ls}}{D_s}
\end{equation}
with the conventional conversion between mass $m$ and the subhalo velocity dispersion $\sigma_v$:
\begin{equation}
\sigma_v = \sqrt{\frac{G m}{2 r_{200}}}.
\end{equation}
For $\left|\delta\right| \rightarrow \infty$, $A$ is given by
\begin{equation}
A(m,\kappa,\gamma)= \frac{3}{2} \pi b^2 \gamma^2 \mu.
\end{equation}

We can use this formalism to compute the optical depth for subhalo lensing as a function of projected distance  $\xi$ from the center of the host halo. At the Einstein radius of the host halo, the magnification diverges to infinity which causes the optical depth at this radius to diverge as well. However, this is not physical since even for point sources the maximum magnification is limited due to the effects of wave optics \citep[e.g.][]{schneider92}. 
Furthermore, the area in the source plane for which a background source would experience very high magnifications is rather small. Of the approximately 1 in 500 quasars which are strongly lensed, only around
1 in $(\mu_{\rm max}-1)^2$ would experience magnifications exceeding $\mu_{\rm max}$. This illustrates the unlikeliness of a galaxy lens producing very high magnifications on a random background quasar (circa one per 
5 million for $\mu_{\rm max}=100$). Of course, any given sample of multiply imaged quasars will be biased toward the high magnification tail since systems with low magnification will fall short of the magnitude limit of the survey. The size of this magnification bias will depend on the quasar luminosity function (QLF) and the magnitude limit of the survey. 
In recent years, there has been a great effort to constrain the QLF using various large surveys producing thousands of QSOs \citep[e.g.][]{boyle00,croom04,richards06a}. However, it has been proven rather difficult to trace the QLF simultaneously to both faint magnitudes and high redshift. Here we use the QLF for $z \sim 3.2$ given in \citet{siana08} covering QSOs at faint-end magnitudes and convolve it with the lensing cross section for a singular isothermal ellipsoid to estimate the effects of magnification bias. For the QLF magnitude limit $r' < 22$, we compute that approximately $0.4\,\%$ of all observed multiply-imaged QSOs will experience magnifications $\mu > 100$ (compared to $\sim 0.01\,\%$ from our simple estimate not taking magnification bias into account). Lowering the magnitude limit to $r' < 20$ roughly doubles the fraction of multiply lensed quasars with $\mu > 100$, thereby demonstrating the importance of the magnitude limit.
Although we expect $z \sim 3.2$ to be a typical value for an observed source redshift, we are also interested in higher redshifts where the effects of magnifaction bias will become even more pronounced.
In \cite{wyithe-loeb}, it was shown that for the Sloan Digital Sky Survey (SDSS) with magnitude limit $i^* < 20$ ($z^* < 20.2$), the fraction of QSOs which are expected to be multiply imaged at $z \sim 4.3$ ($6.0$) might be as high as $\sim 13\,\%$ ($30\,\%$) of which $\sim 5\,\%$ ($10\,\%$) should experience magnifications $\mu>100$.
However, recent results from the SDSS suggest that the QLF bright-end slope at $z > 3$ is getting shallower toward higher redshift \citep{richards06a} and high-resolution $HST$ observations of 157 SDSS QSOs at $4.0<z<5.4$ resulted in the nondetection of strong lensing in these systems \citep{richards06b}. Thus, we put an upper limit $\mu_{\rm max}=100$ for a realistic expected maximum magnification from the host halo and use the \cite{wyithe-loeb} formalism as a conservative estimate of the effect of magnification bias at high redshift.
Currently, the highest redshift quasar with multiple images in the SDSS is at $z_s=3.626$ with its lensing galaxy at $z_l=0.4-0.6$ \citep{inada08} while the most distant multiply-imaged quasar known today is at $z_s=4.5$ with a lensing galaxy at $z_l=0.6$ \citep{mcmahon92}. 

\begin{figure}[t]               
		\plotone{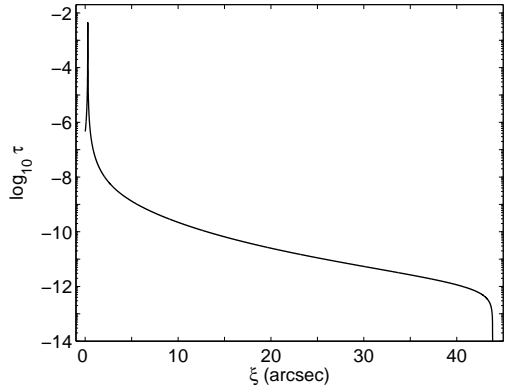}
		\caption{Optical depth for subhalo millilensing of a point source at $z_s$ = 2 as a function of projected radius from the host halo center. Here we assume a host halo at $z_l$ = 0.5 with mass $M = 1.8 \times 10^{12}  \rm{M}_{\odot}$ and subhalos in the mass range $4 \times 10^{6}$ -- 10$^{10} \rm{M}_{\odot}$. The peak has been cut with respect to a maximum magnification factor of 100 at the Einstein radius of the host halo.\label{fig:tau_r}}
\end{figure}

As can be seen in figure \ref{fig:tau_r}, the optical depth for $\mu_{\rm max}=100$ does not exceed a value of 0.005 at any radius. This is contrary to prior claims where the optical depth for subhalo lensing had been estimated to be orders of magnitude higher. \citet{yonehara03} evaluated the quasar millilensing optical depth from SIS subhalos with a slightly tighter mass interval (10$^7$ -- 10$^{10}$ M$_{\odot}$) to be approximately 0.1. If we reconstruct their scenario, we get optical depths below 10$^{-3}$. This discrepancy can be traced to an incorrect subhalo mass function adopted by \citet{yonehara03}.

\begin{figure}[t]
		\plotone{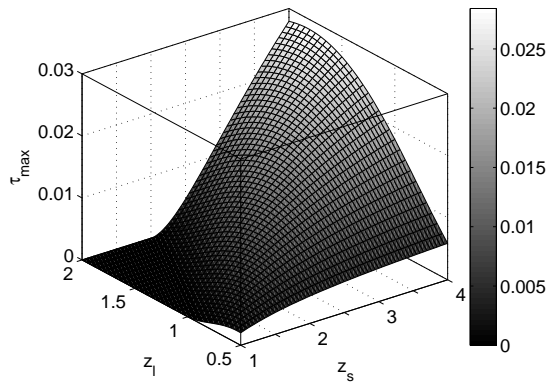}
		\caption{Optical depth for subhalo millilensing corresponding to a maximum magnification factor of 100 at the Einstein radius of the host halo for a point source as a function of lens redshift, $z_l$, and source redshift, $z_s$. We assume a host halo with mass $M = 1.8 \times 10^{12} \rm{M}_{\odot}$ and subhalos in the mass range $4 \times 10^{6}$ -- 10$^{10} \rm{M}_{\odot}$. \label{fig:tau_z}}
\end{figure}

Figure \ref{fig:tau_z} shows the dependence of the optical depth $\tau$ on the lens and source redshifts, $z_l$ and $z_s$ respectively. Here we plot the optical depth at the Einstein radius of the host halo lens with a maximum magnification $\mu_{max}$ = 100. These optical depths are therefore upper limits which will only be valid for quasars favorably aligned with the host halo. Since the optical depth for subhalo lensing increases with both lens and source redshift, observing high-$z$ objects will increase the probability for detecting such an effect considerably. Still, it would require an immense amount of fine-tuning concerning the lens-source alignment in both projected radius and redshift to reach optical depths exceeding 0.025. 

In order to investigate how these results might change taking magnification bias into account, we also estimate the expectation value for the optical depth $<\tau>$ by relaxing the concept of the maximum magnification $\mu_{max}$ and instead using the magnification distribution for multiply-imaged quasars in the SDSS given in figure 6 in \cite{wyithe-loeb}. 
The trend for higher lens and source redshifts to give rise to higher expectation values for the optical depth $<\tau>$ is amplified by the effect of magnification bias. While the expectation value for the optical depth for a system with $z_l = 0.5$ and $z_s \approx 2$ is $<\tau> \sim 10^{-4}$, the corresponding values for $z_s \approx 4$ and $6$ are $<\tau> \sim 10^{-3}$ and $4 \times 10^{-3}$, respectively. Raising the lens redshift for the latter two scenarios to $z_l = 2$, results in an additional increase of the expected optical depth $<\tau>$ by less than one order of magnitude, only reaching values below 0.05. However, as mentioned above, the effect of magnification bias described in \cite{wyithe-loeb} has been shown to be conservative and there are no systems with such redshift combinations known today. Thus, only a dedicated high-$z$ survey, collecting a substantial number ($\gtrsim 100$) of such objects, might challenge our conclusions concerning the bleak prospects for subhalo detection through quasar image splitting in the optical.
\begin{figure}[t]
		\plotone{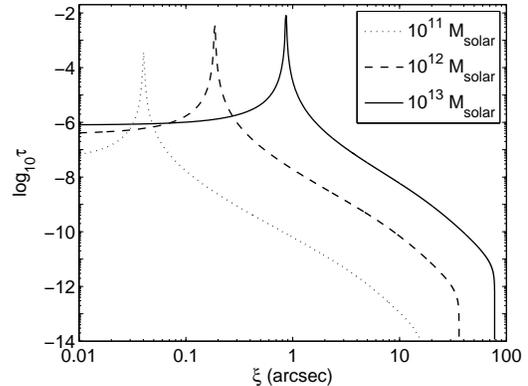}
		\caption{Optical depth for subhalo millilensing of a point source at $z_s$ = 2 as a function of projected radius from the center of a host halo at $z_l$ = 0.5 with mass $M = 10^{11} \rm{M}_{\odot}$ (dotted line), $10^{12} \rm{M}_{\odot}$ (dashed line) and $10^{13} \rm{M}_{\odot}$ (solid line), respectively. For $M = 10^{11} \rm{M}_{\odot}$, $m_{max}$ has been set to $10^9 \rm{M}_{\odot}$. The peak has been cut with respect to a maximum magnification factor of 100 at the Einstein radius of the host halo.\label{fig:tau_M}}
\end{figure}

In figure \ref{fig:tau_M}, we explore the dependence of $\tau$ on the host halo mass. For a point source at $z_s$ = 2 the optical depth for subhalo lensing as a function of projected radius $\xi$ has been computed for a host halo at $z_l$ = 0.5 with mass $M =10^{11}$ M$_{\odot}$, $10^{12}$ M$_{\odot}$ and $10^{13}$ M$_{\odot}$, respectively. 
For a host halo mass of $10^{11}$ M$_{\odot}$ we limit $m_{max}$ to $10^9$ M$_{\odot}$ since the subhalo mass function is not valid for $m > 0.01 M$. Subhalos above this mass limit would also resemble a galaxy-galaxy merger during accretion to its host halo and disturb the system, thereby invalidating previous assumptions. Larger host halos contain a larger fraction in substructure which results in an increased optical depth. However, this effect is not strong enough to substantially raise the probability for detecting subhalo lensing. Even for a host halo mass of $2 \times 10^{13}$ M$_{\odot}$ which is our upper mass limit for a galaxy-sized host halo, the maximum optical depth lies below 0.01.

In summary, we find that the optical depth for strong gravitational lensing of point sources by subhalos is lower than previously assumed, typically well below 0.01. Hence, unless a substantial sample of multiply-imaged quasars would become available, a search for quasar image-splittings by subhalos is unlikely to result in any detections.

\section{Extended sources}
\label{section:ext}
\noindent
If one considers observing the subhalo lensing effects on a quasar at longer wavelength, the quasar can no longer be approximated by a point source. For extended sources, the lensing effects would appear as monopole-like or dipole-like distortion patterns in the surface brightness profile of the source rather than multiple imaging. \citet{inoue05b} have argued that it should be possible to detect these lensing effects from dark matter substructures when observing extended sources resolved at scales smaller than the Einstein radii of the subhalos. Furthermore, it should even be possible to put constraints on the internal density profiles of the lensing subhalos. This technique may already become observationally feasible with ALMA\footnote{www.alma.info} \citep{inoue05a} or future space-VLBI missions like VSOP-2\footnote{www.vsop.isas.ac.jp/vsop2} \citep{inoue05c} and thus constitute a major step forward in the study of dark halo substructures. Here we estimate the probability of subhalo lensing in the radio regime by computing the average number of subhalos that lie within the region of the host halo covering the source. 

In a recent study, \citet{torniainen08} have shown that the typical source size for a quasar ranges from several 10 pc to a few kpc when its turnover frequency falls in the radio regime.

As has been shown by \citet{perrotta02}, the maximum magnification by a galaxy-sized halo that can be achieved for extended sources with an effective radius of 1 -- 10 kpc at redshifts within $z$ = 1 -- 4 falls into the range 10 -- 30. We therefore set a conservative upper limit of 30 for the maximum magnification from the host halo potential. 
\begin{figure}[t]
		\plotone{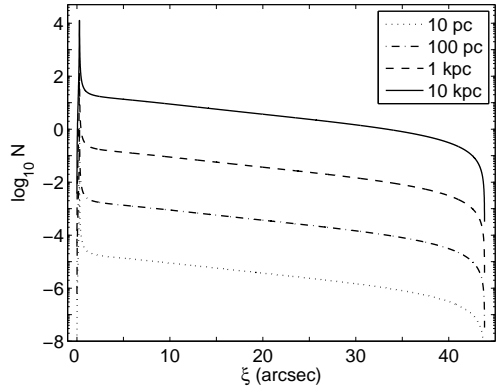}
		\caption{Average number of substructures projected on an extended source at $z_s$ = 2 as a function of projected radius for different source sizes. We assume a host halo at $z_l$ = 0.5 with a mass $M$ = 1.8 $\times$ 10$^{12} \rm{M}_{\odot}$ and subhalos in the mass range 4 $\times$ 10$^{6}$ -- 10$^{10} \rm{M}_{\odot}$. We plot our results for a source with radius $r_s$ = 10 pc (dotted line), 100 pc (dash-dotted line), 1 kpc (dashed line) and 10 kpc (solid line), respectively. We assume a maximum magnification factor of 30 at the Einstein radius of the host halo.\label{fig:N_r}}
\end{figure}

In figure \ref{fig:N_r}, the expected number of subhalos projected on an extended source at $z_s$ = 2 as a function of projected distance from the host halo lens center is shown for several source radii $r_s$, ranging from 10 pc to 10 kpc. This can be compared to the virial radius $R_{200} \approx$ 260 kpc for the host galaxy at $z_l$ = 0.5. It becomes clear that for sufficient source size ($\gtrsim$ 1 kpc) there is a good probability for the source image to be affected by subhalo lensing, not only close to the Einstein radius of the host halo but even at a rather large projected distance from the host halo lens center. For a source with $r_s$ = 1 kpc, one would expect at least one intervening subhalo per 10 observed systems with a maximum projected distance of 10 arcseconds between the foreground galaxy and the source. For $r_s$ = 10 kpc, this number increases to approximately 10 subhalos projected on the source out to a distance of 10 arcseconds from the host galaxy.
This implies that for extended sources not only multiple-image systems but even the much more common singly-imaged quasars should be affected by subhalo lensing. 

Since quasars in the radio are likely to posses intrinsic structure, one has to be careful not to confuse this with a potential lensing signal from subhalo lensing. As internal structures should be mapped on all images of a multiply-imaged source, distortions due to additional lensing of one image on small scales are distinguishable. However, even for singly-imaged quasar one can compare the observed amount of source structure to what would be expected from millilensing by subhalos low-mass field halos. Thus, it opens up the possibility to set upper limits on the abundance and central surface mass density of subhalos using singly-imaged quasars with a foreground galaxy projected within some tens of arcseconds of the source.
\begin{figure}[t]
		\plotone{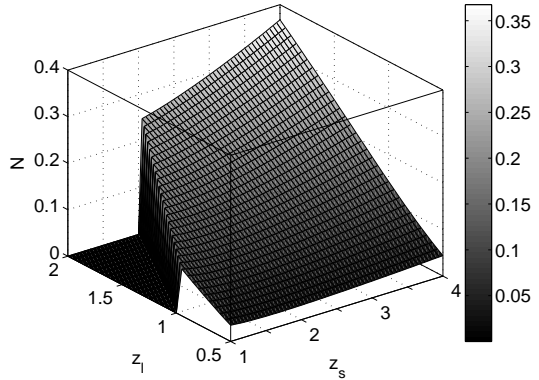}
		\caption{Average number of substructures projected on an extended source with radius $r_s$ = 1 kpc at a projected radius of R$_{200}/2$ (ranging from 22 -- 7 arcseconds for $z_l$ = 0.5 -- 2) as a function of lens redshift, $z_l$, and source redshift, $z_s$. We assume a host halo with mass $M$ = $1.8 \times 10^{12} \rm{M}_{\odot}$ and subhalos in the mass range 4 $\times$ 10$^{6}$ -- 10$^{10} \rm{M}_{\odot}$. \label{fig:N_z}}
\end{figure}

Similar to the case of a point source, we also explore the dependence of these results on the lens and source redshifts, $z_l$ and $z_s$ respectively. Since this scenario seems to be much more feasible than that of a point source, we are not only interested in upper limits but a realistic estimate. The probability for a quasar to be aligned close to the Einstein radius of a foreground galaxy, where its magnification will boost the expected number of intervening subhalo lenses, is low even when considering magnification bias. However, it is expected that there is a large number of galaxy-quasar pairs with a projected separation smaller than the virial radius of the host halo $R_{200}$ (typically up to a few tens of arcseconds). In figure \ref{fig:N_z}, the average number of intervening subhalos at $R_{200}/2$ for a  source size of $r_s$ = 1 kpc is shown as a function of $z_l$ and $z_s$. Also here, the probability for lensing by subhalos increases by up to one order of magnitude for high-$z$ objects.  

\begin{figure}[t]
		\plotone{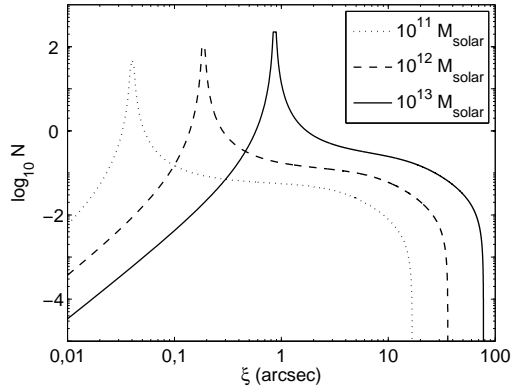}
		\caption{Average number of substructures projected on an extended source with radius $r_s$ = 1 kpc at $z_s$ = 2 as a function of projected radius. Here we assume a host halo at $z_l$ = 0.5 with mass $M$ = $10^{11} \rm{M}_{\odot}$ (dotted line), $10^{12} \rm{M}_{\odot}$ (dashed line) and $10^{13} \rm{M}_{\odot}$ (solid line), respectively. For $M = 10^{11} \rm{M}_{\odot}$, $m_{max}$ has been set to $10^9 \rm{M}_{\odot}$. We assume a maximum magnification factor of 30 at the Einstein radius of the host halo. \label{fig:N_m}}
\end{figure}

Figure \ref{fig:N_m} shows the dependence on the host halo mass of the average number of subhalos intervening with an extended source. We show results for a host halo at $z_l = 0.5$ with mass $M =10^{11}$ M$_{\odot}$, $10^{12}$ M$_{\odot}$ and $10^{13}$ M$_{\odot}$ respectively, and a source with radius $r_s = 1$ kpc at $z_s$ = 2. As in the case for point masses, we limit the maximum subhalo mass $m_{max}$ to $10^9$ M$_{\odot}$ for a host halo mass of $10^{11}$ M$_{\odot}$. As expected, there is an increase with host mass in the average number of subhalos projected on the background source due to the higher fraction in substructure for high-mass host halos. At half the host halo virial radius $R_{200}/2$, this can boost the expected number of subhalos projected on the background source by one order of magnitude over the mass range of galaxy-sized host halos. In addition, high-mass host halos possess larger virial radii $R_{200}$, allowing for larger projected distances between foreground galaxies and background sources.  

All estimates on the number of intervening substructures, $N$, quoted above, however, strongly depend on the minimum subhalo mass, $m_{min}$, set. Since the effects produced by the smallest substructures might not be strong enough to be resolvable with any available or future instrument, there might be an effective minimum subhalo mass above the limit considered here. Implications from this are discussed in $\S$\ref{section:disc}.

\section{Discussion and Conclusions}
\label{section:disc}
\noindent
We have shown that the optical depth $\tau$ for subhalo lensing of point sources is lower than previously predicted. Even for favorable conditions the upper limit on $\tau$ typically lies well below 0.01. The highest optical depths are reached close to the Einstein radius of the host halo where the background source is expected to be multiply-imaged. This implies that for the best probability to observe any effects from subhalo lensing one would have to target multiply-imaged quasar systems at favorable redshifts. Today there are only around 100 of these systems known\footnote{see CASTLES website: http://cfa-www.harvard.edu/castles/} and essentially all measured lens redshifts lie below 1. Adopting the numbers for $\tau$ as shown in figure \ref{fig:tau_z}, where a high host halo magnification of 100 is assumed, we expect that on average 2 images (and no more than 4 images at the 95\% C.L.) out of 100 in this sample could show signs of dark subhalo lensing. Only for high-redshift multiply-imaged quasar systems with lens and source redshifts $z_l \approx 2$ and $z_s > 4$, respectively, the expectation value for the optical depth $<\tau>$ reaches values close to 0.05 when including magnification bias. However, since no such systems are known today, an extensive high-redshift survey would be needed. 

Furthermore, one has to be careful when analysing millilensing signals, since they might not be attributed to halo substructure but low-mass field halos along the line of sight. Although the lensing efficiency for dark matter substructures peaks if they are associated with the host halo, \citet{keeton03} has shown that SIS substructures may be moved in redshift by several tenths and still have a significant lensing effect. 

It is also important to point out that our estimates are valid for substructures with SIS profiles. It has been shown that the Einstein radius of a lens strongly depends on its density profile \citep[e.g.][]{wright00,ez08}. Therefore, adopting a different density profile for these lenses will alter their lensing cross sections and thereby the expected optical depth. N-body simulations based on the CDM paradigm typically predict dark matter halos to have inner density profiles of the form $\rho(r) \propto r^{-\alpha}$ with central density slopes $\alpha \approx$ 1 (e.g. NFW), compared to the SIS with $\alpha$ = 2. The Einstein radius for an NFW type subhalo is several orders of magnitude lower than that for an SIS halo of the same mass. Assuming the subhalos density profiles to follow that proposed by NFW would therefore lower the optical depth for point sources considerably. Even adopting a more favorable density profile with $\alpha =$ 1.5 \citep[][hereafter M99]{moore99} would lower the optical depth computed substantially \citep{ez08}. Recent high-resolution simulations have indicated that substructures might have an inner slope slightly steeper than NFW with significant halo-to-halo variations with $\alpha \approx 1.2$ \citep{diemand08} whereas others favour even shallower inner slopes \citep{navarro04,springel08}. We conclude that it is currently not feasible to use this technique to search for strong lensing signatures of quasars in the optical. 

If one instead targets the radio wavelength regime where quasars appear as extended sources, there is a high probability for subhalo lensing of quasars of sufficient size. For source sizes $r_s \gtrsim$ 1 kpc, this is valid even at rather large projected distance of the source to the host halo center. This allows for a different search strategy than those previously proposed. Instead of only targeting multiply-imaged quasar systems, even quasar-galaxy pairs with a separation of several tens of arcseconds should show effects of strong lensing by substructures in the lens galaxy halo.
\begin{figure}[t]
		\plotone{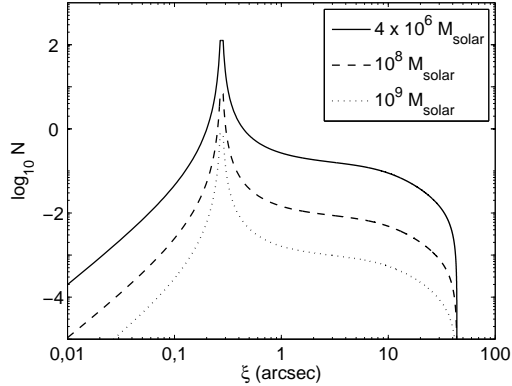}
		\caption{Average number of substructures projected on an extended source with radius $r_s$ = 1 kpc at $z_s$ = 2 as a function of projected radius. For a host halo at $z_l$ = 0.5 with mass $M$ = $1.8 \times 10^{12} \rm{M}_{\odot}$ we vary the minimum subhalo masses with $m_{min}$ = $4 \times 10^6 \rm{M}_{\odot}$ (dotted line), $10^8 \rm{M}_{\odot}$ (dashed line) and $10^9 \rm{M}_{\odot}$ (solid line), respectively. We assume a maximum magnification factor of 30 at the Einstein radius of the host halo. \label{fig:N_m_max}}
\end{figure}

However, these effects will strongly depend on the subhalo mass and density profiles. In paper I, we have shown that the image separation caused by substructures in the mass range $\lesssim 10^{10}$ M$_{\odot}$ will only be resolvable if one can assume a steep inner density profile similar to SIS and M99. Even for these favorable density profiles, there will be a minimum mass for which the image separation drops below the resolution of any present or planned observational facilities. For instance, for the currently available EVN\footnote{www.evlbi.org}, the minimum subhalo mass which could be resolved under optimal conditions is approximately $4 \times 10^6$ M$_{\odot}$ in case of an SIS profile but increases to $3 \times 10^7$ M$_{\odot}$ for an M99 profile. Figure \ref{fig:N_m_max} shows how the expected number of intervening substructures $N$ depends on the minimum subhalo mass $m_{min}$. For a host halo of mass $M$ = $1.8 \times 10^{12}$ M$_{\odot}$ at $z_l$ = 0.5 and a source of size $r_s$ = 1 kpc at $z_s$ = 2, we compute $N$ as a function of projected radius from the host halo center for minimum subhalo mass $m_{min}$ = $4 \times 10^6$ M$_{\odot}$, $10^8$ M$_{\odot}$ and $10^9$ M$_{\odot}$, respectively. Since the subhalo mass function predicts about equal mass within each logarithmic mass bin, most subhalos will be of low mass ($n(m) \propto m^{-0.9}$). Thus, the expected number of substructures projected on the source is very sensitive to the minimum subhalo mass that can be resolved. 

Therefore, angular resolution will be crucial when attempting to detect CDM substructure via its lensing effects on background quasars and submilliarcsecond-resolution facilities will be required. We assess that the upcoming VSOP-2 satellite may provide the best prospects for such a detection. However, one must be careful not to confuse internal structures found in quasars observed at radio wavelengths with subhalo lensing signals. Quasars already macrolensed on arcsecond scales can be used to test that it is possible to distinguish between the two, since internal structures should be mapped in all of the macro-images while subhalo lensing will only affect one of the images. Therefore, such system will be ideal to train this technique of identifying the strong lensing signal from dark matter substructures. However, even single-imaged systems with rather large projected lens-source distance could be used to set upper limits on the abundance and nature of subhalos depending on the amount of distortion signals detected. Taking the above into account, one may be able to use this technique to put constraints on dark matter subhalos predicted by simulations in the near future.

\acknowledgments
We are grateful to the referee for the constructive comments, which helped to improve the paper.
TR acknowledges support from the HEAC Centre funded by the Swedish Research Council. EZ acknowledges research grants from the Academy of Finland, the Swedish Research Council and the Royal Swedish Academy of Sciences. EM acknowledges support from the Swedish Research Council and the Royal Swedish Academy of Sciences. KW acknowledges support from the Jenny and Antti Wihuri foundation.

\clearpage

\end{document}